\documentclass[prl,amsmath,amssymb,superscriptaddress,twocolumn]{revtex4}
\usepackage{graphicx}
\usepackage{float}
\usepackage{amsthm,amsfonts}

\newcommand{\ket}[1]{\mathop{\left|#1\right>}\nolimits}            
\newcommand{\bra}[1]{\mathop{\left<#1\,\right|}\nolimits}         

\def\e{\epsilon}

\def\e{\mathrm{e}}

\begin{document}
\title{Synchronization and estimation of gravity-induced time difference for quantum clocks}

\author{Jieci Wang$^{1}$\footnote{jcwang@hunnu.edu.cn}, Tonghua Liu$^{1,2}$\footnote{295609904@qq.com},  Jiliang Jing$^{1}$\footnote{jljing@hunnu.edu.cn}, and Songbai Chen$^{1}$ \footnote{csb3752@hunnu.edu.cn}}
\affiliation{$^1$ Department of Physics,
 Hunan Normal University, Changsha, Hunan 410081, China\\
 $^2$ Department of Astronomy, Beijing Normal University, Beijing 100875, China}.

\begin{abstract}

It has recently been reported [\textit{PNAS} \textbf{114}, 2303 (2017)] that, under an operational definition of time,   quantum clocks would
get entangled through gravitational effects.
Here we study an alternative scenario: the clocks have different masses and energy gaps, which would produce  time difference via gravitational interaction. The proposal of quantum clock  synchronization  for the gravity-induced time difference is discussed. We illustrate how the stability of measurement probability in the quantum clock synchronization proposal is influenced by the gravitational interaction induced by the clock themselves. It is found that the precision of clock synchronization  depends on the energy gaps of the clocks and the improvement of precision in quantum metrology is in fact an indicator of entanglement generation. We also present the quantum enhanced  estimation of time difference and find that the  quantum Fisher information is very sensitive to the distance between the clocks.

\end{abstract}

\maketitle

\section{ Introduction}
 The present quantum clocks are conventionally considered as ideal objects since the flow of time isn't
affect by nearby clocks \cite{buzek}.
This assumption is artificial because the clocks are not treated  as  physical entities. However, according to general relativity, the picture of spacetime  assigns  each quantum clock a worldline  \cite{igor,taylor}. Therefore,  gravitational effects of the clocks cannot be ignored and the flow of time according to one clock is indeed influenced by the presence of clocks along nearby worldlines  \cite{busch,brown}. Furthermore,  realistic quantum clocks, which consist of microscopic particles such as atoms, should obey the theory of quantum mechanics \cite{clock4, rovelli}.
 Recently,  Castro-Ruiz \emph{et al.} found that  the clocks necessarily get entangled through gravitational interactions if time is defined operationally \cite{ECRG}.   Marletto \emph{et al.}  \cite{CMVV} proposed a  scheme to detect the generated  entanglement between two masses via gravitational interactions. In addition, quantum entanglement is found to be significant at energy scales which exist naturally in sub-atomic particle bound states \cite{AYTY}.

 As showed in \cite{ECRG}, the clock's precision is inversely proportional to the energy difference between the eigenstates and each energy eigenstate of the clock corresponds to  a particular  strength of gravitational field. According to general relativity, different energy gaps lead to different time dilation  in  the clocks. Therefore, the time records may be different in the clocks because they experience different mass-induced gravitational field background. In other words, the quantum clocks would become unsynchronized if the masses and  energy gaps of the clocks are not identical.
From this perspective  the study of clock synchronization  for gravity-induced time difference will give some new insights in relativistic quantum information, which is an interdiscipline of  quantum physics and general relativity.

In this paper we study  how to synchronize gravity-induced  time difference via a quantum clock synchronization (QCS) proposal and the  estimation of the gravity-induced time difference. The QCS  plays
an important role in modern society and scientific research
such as fundamental physics, astronomy and navigation \cite{sysapp1,sysapp2}. In addition, the synchronization schemes based
 on quantum mechanics can gain significant improvements in precision over their classical counterparts \cite{RJDA,IC,VGSL,qsyn2,qsyn3,qsyn4,qsyn5,qsyn6,qsyn7,MKPP,RBAI,JWZT,LZJJ}.
Unlike previous QCS protocols, our  quantum clocks are two-level  particles in the superposition of energy eigenstates and gravitational effects only originate from the clock themselves.  We present how  the precision of QCS is affected by  the energy gaps of the clocks and the distance between the clocks.   We also employ the method of quantum metrology to  enhance the  estimation of the gravity-induced time difference. It is found that the precision for the estimation of time difference depends on energy gaps.  In addition, the improvement of precision in quantum metrology is found to be an indicator for the  generation of quantum correlation.

The outline of the paper is organized as follows:
In Sec. II, we introduce the evolution of two clocks under  gravitational interactions.
In Sec. III, we study the two-party QCS for the gravity-induced time difference. The quantum enhanced  estimation of time difference between two clocks is studied in Sec. IV. We make a conclusion in the last section.

\section{ The evolution of quantum clock under the gravitational interaction}
A system with a superposition of energy eigenstates can be considered as a
 quantum clock. The simplest case of a  clock is
 a  two-level  particle which  follows an   approximately static semiclassical trajectory, which has approximately  zero  momentum with respect to the far away observer. The clocks are required to be point like because only in this case one can  assign each quantum clock a world-line.  In addition, the clock (particle)'s spacetime has a non-fixed metric background when the  quantum mechanical superposition of energy eigenstates is considered \cite{ECRG,peskin}.

The model of the present paper is based on the following assumptions \cite{ECRG}:  (i) the general relativistic mass-energy equivalence which implies gravitational interaction between the clocks; (ii)  the quantum mechanical superposition of energy eigenstates which leads to a non-fixed metric background.    The whole mass-energy contribution to the gravitational field can be interpreted as a sum of static mass and internal energy (dynamical mass) corresponding to the energy of the internal degrees of freedom \cite{ECRG}. It is worthy noting that the gravitational effect of space-time background is not considered in the present model. In fact,  the dynamical mass is of  relativistic nature, which arises from the gravitational  interaction between the clocks. The clocks would become unsynchronized because the masses and  energy gaps of the clocks are not identical. Therefore, precision of quantum clock synchronization is influenced by the gravitational interaction induced by the clock themselves.

We implement the mass energy equivalence by considering the composite particles are emerged from the interaction of two scalar fields $\varphi_1$ and $\varphi_2$ \cite{ECRG,peskin}.
The Lagrangian density of the scalar field is
\begin{equation}
\mathcal{L} = - \frac{1}{2}\sqrt{-g}\left(\sum_Ag^{\mu\nu}\left(\partial_{\mu}\varphi_A\right)\left(\partial_{\nu}\varphi_A\right) + \sum_{AB} M^2_{AB}\varphi_A\varphi_B\right),\end{equation}
where $g$ is the determinant of the spacetime metric $g_{\mu\nu}$, and $M_{AB}$ is the symmetric matrix in which the fields $\varphi_1$ and $\varphi_2$ are coupled \cite{ECRG,peskin}. For the sake of simplicity, we employ natural units $(c=\hbar=1)$ in this part. The eigenvalues of $M_{AB}$ can be denoted  by $m$ and $m+\Delta E$.    In general relativity, there is fundamentally no difference between mass and interaction energy. Therefore,  the distinction between the static mass and the  dynamical mass depends on the energy scale with which the system is probed. In this sense, the matrix $M_{AB}$ includes both the static mass ($m$) contribution and the internal energy (or dynamical mass  with eigenvalue $\Delta E$) contribution.
In the weak field limit \cite{ECRG,peskin}, the non-fixed metric background of the particle reads  $\mathrm{d}s^2 = -(1+2\Phi(\mathbf{x}))\mathrm{d}t^2+\mathbf{\mathrm{d}x \cdot \mathrm{d}x}$, where  $\Phi$ is the gravitational potential. Then the Hamiltonian is obtained  via the Legendre transformation, which is
\begin{equation}
\label{fieldhamiltonian}
H = \frac{1}{2} \int \mathrm{d}^3\mathbf{x}\left(\sum_A \left(\pi_A^2 +(\nabla\varphi_A)^2 \right) + \sum_{AB}M^2_{AB}\varphi_A \varphi_B\right)\mathbf{\Phi},
\end{equation}
where $\mathbf{\Phi}=\left(1+\Phi(\mathbf{x})\right)$, and $\pi_A = \dot{\varphi_A} $ is the canonical conjugate momentum respects to $\varphi_A$. 

To quantise the field, we diagonalize the matrix $M_{AB}$ and  write the Hamiltonian as $H = \frac{1}{2} \sum_A \int \mathrm{d}^3\mathbf{x}\mathbf{\Phi}\left(p_A^2 +(\nabla\Psi_A)^2  + \mu^2_{A}\Psi_A^2\right)$,
where $\Psi_A = \sum_B C_{AB} \varphi_B$,  $p_A$ is the momentum conjugate to $\Psi_A$, and $C_{AB}$ is the matrix diagonalizing $M_{AB}$.
By empolying the slow velocity approximation of the fields and Fourier-expanding $\Psi$ and $p_A$, we obtain
$
\Psi_A(\mathbf{x})\approx  \frac{1}{\sqrt{2\mu_a}}\left(\chi_A(\mathbf{x})+\chi_A^\dagger(\mathbf{x})\right)$ and
$p_A(\mathbf{x}) \approx  \, \mathrm{i}\,\sqrt{\frac{\mu_a}{2}}\left(\chi_A(\mathbf{x})-\chi_A^\dagger(\mathbf{x})\right)$,
where $\chi_A(\mathbf{x}) = (2\pi)^{-3}\int \mathrm{d}^3\mathbf{k}\mathrm{e}^{\mathrm{i}(\mu_At-\mathbf{k}\cdot \mathbf{x})}b_{A,\mathbf{k}}$. The gravitational potential $\Phi$ obeys the Poisson equation $\nabla^2\Phi = -4\pi \rho$, where $\rho $ is the energy density of the matter field.  Solving the Poisson equation under the gravitational potential
$
\Phi(\mathbf{x}) = -G \int \mathrm{d}^3\mathbf{x^\prime}\frac{\rho(\mathbf{x^\prime})}{\vert \mathbf{x}-\mathbf{x^\prime}\vert}
$
and inserting the solution into (\ref{fieldhamiltonian}),  one obtains the normal-ordered Hamiltonian for the clock particle
\begin{eqnarray}
\label{matrixhamiltonian}
\nonumber H&=& \sum_{AB}\int \mathrm{d}^3\mathbf{x} \left(M_{AB} \phi^\dagger_A \phi_B- \frac{1}{2} M^{-1}_{AB} \phi^\dagger_A\nabla^2\phi_B\right) \\ &&-G	\sum_{ABCD}\int  \mathrm{d}^3\mathbf{x}{d}^3\mathbf{x^\prime} \frac{(M_{AB}\phi^\dagger_A\phi_B)(M_{CD} \phi^\dagger_C \phi_D)}{\vert \mathbf{x}-\mathbf{x^\prime}\vert},\nonumber
\end{eqnarray}
where $\phi_A = \sum_B C^{-1}_{AB}\chi_{B}$ and
 $ \rho(\mathbf{x}) = \sum_{AB} M_{AB} \phi^\dagger_A(\mathbf{x}) \phi_B(\mathbf{x})$ is the mass-energy density of the field. However, it is not a well-defined operator  since it involves the product of two field operators at the same point which induces divergencies. This requires  a suitable regularization procedure and leads to a renormalisation of the mass [16], which is $ \rho_{reg}(\mathbf{x}) = \sum_{AB} \int \mathrm{d}^3\mathbf{x^\prime} f_{\delta} (\mathbf{x}-\mathbf{x^\prime})M_{AB} \phi^\dagger_A(\mathbf{x^\prime}) \phi_B (\mathbf{x^\prime})$. In this expression $ f_{\delta}$ is a normalised  positive function, which reduces to $f_{\delta_r}(\mathbf{x})\longrightarrow\delta^3(\mathbf{x})$ when the regularization parameter  $\delta_r\rightarrow0$.

To obtain the evolution of the two-particle state in the Fock space, we calculate the matrix element $\bra{\xi^{(1)},\eta^{(1)}} H \ket{\xi^{(2)},\eta^{(2)}}$, where $\ket{\xi^{(i)},\eta^{(i)}} = 2^{-1/2}\sum_{AB}\int \mathrm{d}^3\mathbf{x}{d}^3\mathbf{x^\prime} \xi_A (\mathbf{x}) \eta_B(\mathbf{x^\prime})\phi^\dagger_A(\mathbf{x})\phi^\dagger_B(\mathbf{x^\prime})\ket{0}$ is a two-particle state for $i = 1,2$, and $\ket{0}$ denotes the vacuo of the field. Then the two particle Hamiltonian is found to be
\begin{eqnarray}
\label{twoparticlehamiltonian}
\nonumber \hat{H} &=& \hat{M}_{r}\otimes I+I\otimes\hat{M}_{ren} + \frac{1}{2}\hat{M}^{-1}\hat{\mathbf{p}}^2\otimes I\\
&&+ \frac{1}{2}\hat{M}^{-1}I\otimes\hat{\mathbf{p}}^2 -G\frac{\hat{M}\otimes\hat{M}}{\vert \hat{\mathbf{x}}\otimes I-I \otimes\hat{\mathbf{x}} \vert},	
\end{eqnarray}
where $\hat{M}_{r} = \hat{M}- (\pi\delta^2)^{-1/2} G \hat{M}^2 $ is the renormalized mass matrix and $\bra{\xi}\hat{M}\ket{\eta} = \sum_{AB}\int \mathrm{d}^3\mathbf{x} \bar{\xi}_A(\mathbf{x})M_{AB}\eta_B(\mathbf{x})$.  By projecting it in the corresponding subspace, one can obtain the Hamiltonian for an arbitrary number of particles.

The whole mass-energy contribution to the gravitational field can be interpreted as a sum of static mass and internal energy (dynamical mass), which corresponds to the energy of the internal degrees of freedom  $M\rightarrow \hat{H}_i/c^2$ with $i=A, B$ \cite{ECRG}. According to relativistic mass-energy
equivalence, the dynamical mass is of a purely relativistic nature, and arises from the interaction between the clocks.
In the low velocity limit,   the kinetic part  since the static mass is negligibly small as compared to the dynamical mass. Thus,
the Eq. (\ref{twoparticlehamiltonian}) for the two-clock  system reduce to \cite{ECRG,peskin}
\begin{equation}\label{HMTL}
\hat{H}_{tot}=\hat{H}_A+\hat{H}_B-G\frac{\hat{H}_A\hat{H}_B}{\mathbf{x}_A-\mathbf{x}_B}.
\end{equation}

We assume that the initial state of the clocks  is an uncorrelated state and the clocks are fixed at a static background
\begin{equation}\label{initial}
\ket{\psi_{in}}_{AB} =\ket{\psi}_A{\otimes}\ket{\psi}_B,
\end{equation}
where $\ket{\psi}_{A(B)}=\frac{1}{\sqrt{2}}(\ket{0}_{A(B)} + \ket{1}_{A(B)})$.
Employing the Hamiltonian given by Eq. (\ref{HMTL}) and releasing the natural units, the evolution of the whole system at time $t$ according to the far away observer C  is found to be
\begin{eqnarray} \label{final}
\nonumber\ket{\psi_{f}(t)}_{AB}
\nonumber&=&\frac{1}{2}\big(\ket{0_A}\ket{0_B}+\e^{-\frac{i}{\hbar}
\Delta E_2t}\ket{0_A}\ket{1_B}\\ && \nonumber+\e^{-\frac{i}{\hbar}\Delta E_1t}\ket{1_A}\e^{-\frac{i}{\hbar}\Delta E_2(1-\frac{G\Delta E_1}{ c^4x})t}\ket{1_B} \\ && +\e^{-\frac{i}{\hbar}
\Delta E_1t}\ket{1_A}\ket{0_B} \big),
\end{eqnarray}
which shows that the clocks get entangled through gravitational interaction.  If the clocks have identical mass and energy gaps, the final state Eq. (\ref{final}) for the clocks is consistent with Eq. (5) in \cite{ECRG}.
  In the previous equation, the following facts are used:  $\hat{H}_A\ket{0}_A= E_0\ket{0}_A, \hat{H}_A\ket{1}_A= E_1\ket{1}_A$ and $\hat{H}_B\ket{0}_B= E_0\ket{0}_B, \hat{H}_B\ket{1}_B= E_2\ket{1}_B$. For convenience, we have put $E_0=0$,
therefore the energy gaps can be defined as  $\Delta E_1=E_1-E_0=E_1$ and $\Delta E_2=E_2-E_0=E_2$. In fact, Eq. (\ref{final}) describes the evolution of clock $A$ and clock $B$ under the time coordinate $t$ of the observer $C$.  Since the entire system is  observed by the far away observer C,   the time coordinate $t$ of the observer $C$ is  selected as the time-line  \cite{ECRG}.

It is worthy to mention that the picture studied here is different from  \cite{ECRG} because the energy gaps of the clocks are not identical to each other, which leads to different dynamic mass and different metric background.
That is to say, the time of each clock is  changed due to the unfixed spacetime metric  arising from nearby clocks.   Based  on the assumptions that both  hold in this situation, we can safely conclude that the evolution of clock A is different from clock B at excited state due to gravitational interaction. In other words,  different energy gaps for the clocks leads to different time dilation. Therefore, clock $A$ does not synchronize with clock $B$ anymore and clock synchronization is required in this situation.

\section{ Quantum clock synchronization for gravity-induced time difference}
In this section, we present the QCS protocol between clock $A$ and clock $B$ when the initial state of the clocks is given by Eq. (\ref{initial}).
The main task of clock synchronization is to determine the time difference between two spatially separated clocks.
 We assume that clock $A$ is the standard clock,  while clock $B$ need to be synchronized due to the gravity-induced  time dilation. In addition,  Alice and Bob arrange a  starting  time $\tau$ of their measurements at the beginning of the evolution. Since they do not have a pair of  synchronized
clocks to start with, they have to start the measurement at the arranged proper time $\tau_i  (i=A,B)$  relative to the time reading of their local
clocks, which are different due to the gravitational interaction.
To synchronize the  clocks, we adopt the dual basis $\ket{+}=\frac{1}{\sqrt{2}}(\ket{0}+\ket{1})$ and
$\ket{-}=\frac{1}{\sqrt{2}}(\ket{0}-\ket{1})$ as the measurement basis, which can be obtained from $\ket{0}$ and $\ket{1}$ through the Hadamard transformation. Then the total system has following form
\begin{eqnarray}
\rho_{AB}(t) = \frac{1}{16}\left(
                  \begin{array}{cccc}
                     |\beta|^2& \beta \alpha^* & \beta \eta^* &  \beta \gamma^*\\

                     \alpha \beta^* & |\alpha|^2  & \alpha\eta^* & \alpha \gamma^* \\

                     \eta \beta^* &   \eta\alpha^* &  |\eta|^2 &   \eta \gamma^*\\

                      \gamma \beta^*& \gamma \alpha^* &\gamma \eta^* & |\gamma|^2 \\
                  \end{array}
                \right)
,\label{clockac}
\end{eqnarray}
in the ${\ket{++},\ket{+-},\ket{-+},\ket{--}}$ basis, where
 \begin{eqnarray}
 \nonumber \alpha&=&(1+\e^{-\frac{i}{\hbar}\Delta E_1t}-\e^{-\frac{i}{\hbar}\Delta E_2t}-\e^{-\frac{i}{\hbar}\Delta E_1t}\e^{-\frac{i}{\hbar}\Delta E_2^{'}t}),\\
 \nonumber\beta&=&(1+\e^{-\frac{i}{\hbar}\Delta E_1t}+\e^{-\frac{i}{\hbar}\Delta E_2t}+\e^{-\frac{i}{\hbar}\Delta E_1t}\e^{-\frac{i}{\hbar}\Delta E_2^{'}t}),\\
 \nonumber\eta&=&(1-\e^{-\frac{i}{\hbar}\Delta E_1t}+\e^{-\frac{i}{\hbar}\Delta E_2t}-\e^{-\frac{i}{\hbar}\Delta E_1t}\e^{-\frac{i}{\hbar}\Delta E_2^{'}t}), \\
 \nonumber \gamma&=&(1-\e^{-\frac{i}{\hbar}\Delta E_1t}-\e^{-\frac{i}{\hbar}\Delta E_2t}+\e^{-\frac{i}{\hbar}\Delta E_1t}\e^{-\frac{i}{\hbar}\Delta E_2^{'}t}),
 \end{eqnarray} and  \begin{eqnarray}
 \nonumber\Delta E_2^{'}=\Delta E_2(1-\frac{G\Delta E_1}{ c^4x}).\end{eqnarray}
From the viewpoint of  the observer $C$,  the  entire system evolves   under the time coordinate $t$. To find out  the time difference, we assume that the observer $C$ observes
 the entire system far from the clocks, while  Alice and Bob measure their own local quantum system. After measuring each qubit   at proper time $\tau_A=\tau$, Alice publishes the results of her measurement.   It's worth noting that clock $A$ and clock $B$ measure the system according to their own proper time.  However,  Bob's proper time  is changed by the gravitational effects induced by the mass of Alices' clock since the energy gap of the clocks is not identical to each other.  In other words, it would be a  time difference $\tau_B-\tau_A=\delta$ between the proper time $\tau_A$ and the proper time $\tau_B$. After Alice's  measurement,  Bob's state  immediately collapses to

\begin{eqnarray}
\rho_{B}(\tau_B=\tau) = \frac{1}{12+4\cos(\zeta\delta)}\left(
                  \begin{array}{cc}
                      |\varsigma|^2&\varsigma \kappa^*  \\

                    \kappa \varsigma^* & |\kappa|^2   \\

                  \end{array}
                \right)
,
\end{eqnarray}
where $\varsigma=(2+\e^{-\frac{i}{\hbar}\Delta E_2\delta}+\e^{-\frac{i}{\hbar}\Delta E_2^{'}\delta})$, $\kappa=(2-\e^{-\frac{i}{\hbar}\Delta E_2\delta}-\e^{-\frac{i}{\hbar}\Delta E_2^{'}\delta})$
 and $\zeta=\frac{G\Delta E_1\Delta E_2}{ c^4x\hbar}$ normalizes the final density
operator.
Therefore, the probability for Bob measuring $|+\rangle$ or $|-\rangle$ at time $\tau_{B}=\tau$ are
\begin{eqnarray}\label{prob}
 P_B(\ket{\pm})&=&\frac{1}{2}\pm\zeta_1\bigg(\cos(\frac{\Delta E_2\delta}{\hbar})
+\cos(\frac{\Delta E_2^{'}\delta}{\hbar}) \bigg),
\end{eqnarray}
where $\zeta_1=\frac{1}{3+\cos(\zeta\delta)}$. This allows us to estimate the time difference $\delta$ between the clocks after the gravitational interaction.
Then the information of   $\delta$  is exposed by the observable probabilities and Bob can adjust his clock accordingly.

\begin{figure}[tbp]
\centering
\includegraphics[height=2.5in, width=2.5in]{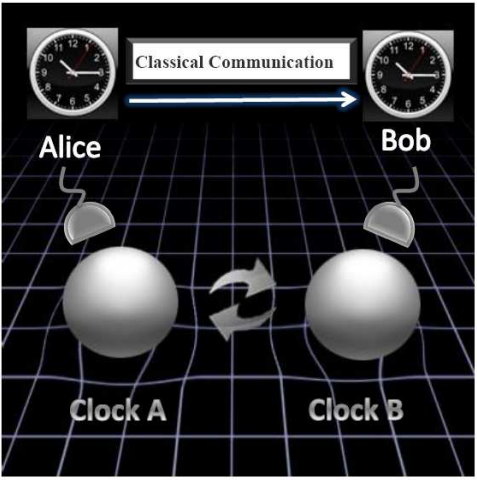}
\caption{(Color online) The scheme for quantum clock synchronization. Proper time in the clocks are influenced by the gravitational interaction induced by the clock themselves. Alice and Bob measure their own local quantum systems at the prearranged proper time $\tau$. Alice measures her subsystem  at proper time $\tau_A=\tau$ and  publishes the result of the measurement by classical communication. Bob adjusts his clock based on the  probabilities of his local measurements.
}\label{f1}
\end{figure}

Considering the interaction distance
of quantum effect in the gravitational field is exceptionally small, we can employ the dimensionless  Planck units to develope  a quantitative study. To this end we define $\delta_p=\delta/t_p$, $\varepsilon=\Delta E/E_p$, $\xi=x/l_p$, where $l_p=\sqrt{\hbar G/c^3}$ is the Planck length, $t_p=l_p/c$ is the Planck time and $E_p=\hbar/t_p$ is the Planck energy. Without loss of generality, the  probabilities in Eq. (\ref {prob}) can be rewritten as
\begin{eqnarray}\label{pro}
 P_B(\ket{\pm})=\frac{1}{2}\pm\zeta_2[\cos(\varepsilon_2\delta_p)
+\cos((\varepsilon_2-\zeta^{'})\delta_p) ],
\end{eqnarray}
where $\zeta_2=(3+\cos(\frac{\varepsilon_1\varepsilon_2}{\xi}\delta_p))^{-1}$.
It is not difficult to recognize that the term $\cos((\varepsilon_2-\zeta^{'})\delta_p)$ is induced by the gravitational interaction. In the limit of  $\zeta^{'}\rightarrow0$, which corresponds to the limit of the distance between the  clocks is much longer than the Planck length, the measurement probabilities for clock $B$ became $P_B(\ket{\pm})=\frac{1}{2}\pm\frac{1}{2}\cos(\varepsilon_2\delta_p)$.  This result is consistent with the measurement probabilities in flat spacetime \cite{RJDA,MKPP,RBAI}.

In Fig. (1), we show the measurement probability $P_B$ with respect to variations of energy gap $\varepsilon_1$ for different distances.
The measurement probability $P_B$ are found to be periodic oscillation with  increasing energy gap of the clock $A$. That is to say, the stability of the QCS proposal is influenced by the gravitational interaction induced by the clock themselves.  In addition, the period of oscillation increases with increasing distance between Alice and Bob, which means that the magnitude of the gravitational field reduced by the mass of clock A inevitably influences the performance of  QCS. It is not difficult to infer that if clock $A$ is very far away from clock $B$, the measurement probability would tend to a steady value. This indicates that the probability of clock synchronization is not affected by the negligible gravitational interaction in this case.

\begin{figure}[tbp]
\centering
\includegraphics[height=2.0in, width=3.0in]{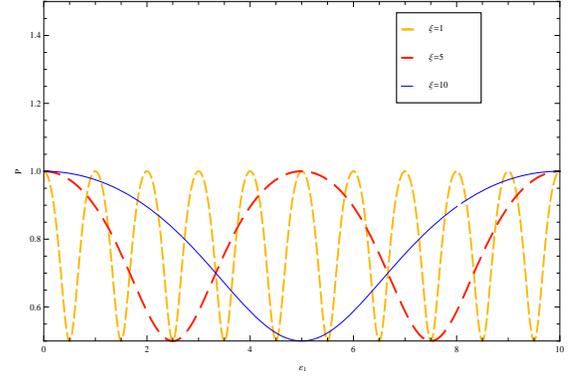}
\caption{(Color online) The measurement probability as a function of the energy gap $\varepsilon_1$ with different distances $\xi=1$ (orange dotted line), $\xi=2$ (red dotted line) and $\xi=10$ (blue solid line) between two clocks, respectively.
 The energy gap and the time difference parameter $\delta$
are fixed as $\varepsilon_2=10$ and $\varepsilon_2 \delta_p=2\pi$.
}\label{f3}
\end{figure}

\section{Quantum metrology and the estimation of the time difference}
 Quantum metrology aims  to estimate a parameter with
higher precision than classical approaches by using quantum approaches  \cite{advances}. For a given measurement scheme, the ultimate limit on the variance of the errors for the parameter $\theta$ is bounded by the Cram\'{e}r-Rao inequality \cite{Cramer:Methods1946,APM} $Var(\theta) \geq  [n\mathcal{F}_{Q}(\theta)]^{-1}$,
where $n$ is the number of repeated measurements and $\mathcal{F}_{Q}(\theta)\geq \mathcal{F}_{\xi}(\theta)$ is the quantum Fisher information.
Recently, the adaptation of quantum metrology to improve probing technologies of relativistic effects has been studied in different contexts \cite{aspachs,HoslerKok2013, mostrecent,RQM,RQM2,wangtian2}. In this section we  study the quantum parameter estimation for the time difference. To this end, we calculate the quantum Fisher information of the final state of Bob
\begin{eqnarray}
\nonumber\ket{\psi_f}_{B}&=&\frac{1}{\sqrt{12+4\cos(\zeta\delta)}}\bigg((2+\e^{-\frac{i}{\hbar}\Delta E_2\delta}+\e^{-\frac{i}{\hbar}\Delta E_2^{'}\delta})\ket{+}\\
&+&(2-\e^{-\frac{i}{\hbar}\Delta E_2\delta}-\e^{-\frac{i}{\hbar}\Delta E_2^{'}\delta})\ket{-}\bigg),
\end{eqnarray}
which is the probe state for quantum metrology.
Basing on the measurement results of $\ket{\psi_f}_{B}$, one can obtain an estimate of $\delta$ and denote it as $\dot{\delta}$. The estimation precision of $\delta$ can be described by $\Delta \delta$, where
$\Delta \delta=\sqrt{\langle(\dot{\delta}-\delta)^2\rangle}$ and the average is taken over all the possible measurement results.  We further note that for the single parameter case,
the equality can be obtained, such that $\Delta\delta=1/\sqrt{\mu F_Q}$. Since we are only interested in the quantum enhancement of the precision, we will set $\mu=1$, so $\Delta\delta=1/\sqrt{ F_Q}$. The quantum Fisher information of the final state is found to be
$
F_Q=\frac{(2\cos(\zeta^{'}\delta)-1)[(2\varepsilon_2^2-2\varepsilon_2\zeta^{'})\cos(\zeta^{'}\delta)+
\varepsilon_2^2+(\varepsilon_2-\zeta^{'})^2)]}{3+\cos(\zeta^{'}\delta)}.
$
When the clocks are  very  far away from each other, we have $\zeta^{'}\rightarrow0$.  Taking the Taylor expansion and ignoring the  second order term, $F_Q$ has the following form
$
F_Q\simeq\frac{(2\varepsilon_2-\zeta^{'})^2}{4}$,
which leads to
$\Delta \delta\simeq\frac{2}{(2\varepsilon_2-\zeta^{'})}$.
Then we can see that  the precision of quantum parameter estimation on the time difference is consistent with the results in the flat spacetime \cite{JDYY} when the clock $A$ is far enough away from clock $B$.

\begin{figure}[tbp]
\centering
\includegraphics[height=2.0in, width=3.0in]{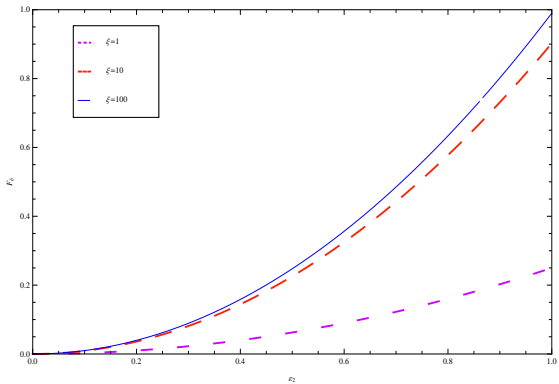}
\caption{(Color online) The QFI for the estimation of time difference as a function of the energy gap of clock $B$ for different distances $\xi=1$ (violet dotted line), $\xi=10$ (red dotted line) and $\xi=100$ (blue solid line) between the clocks. The energy gap of clock $A$ are fixed as $\varepsilon_1=10$
}\label{Fig2}
\end{figure}

In Fig. (\ref{Fig2}), we plot the quantum Fisher information of the time difference as a function of the  energy gap $\varepsilon_2$ for different interaction distances.  It is shown  that the quantum Fisher information increase with the growth of the energy gap $\varepsilon_2$, which means that the precision for estimating time difference depends to energy gap $\varepsilon_2$ of the clocks. It was found in \cite{ECRG} that quantum entanglement is generated by the gravitational interaction induced by the masses of the clocks. Therefore, the improvement of precision in quantum metrology is in fact an indicator of entanglement generation  in a relativistic setting \cite{sunwang}. On the other hand, the increase of quantum Fisher information verifies the fact that the precision of quantum metrology can be enhanced by quantum resources such as entanglement.  In addition, the  quantum Fisher information is found to be very discriminable for different distances  between two clocks. It is worthy noting that the quantum Fisher information doesn't have  significant changes for larger distances, which indicates that effects of gravitational interaction induced by the massed of the clocks is very sensitive to the distance.

\section{ Conclusion} In conclusion, we have studied the time difference induced by gravitational interaction and the protocol of QCS for two spatially separated clocks with different energy gaps.
It is shown that the stability of the QCS proposal is influenced by the gravitational interaction induced by the clock themselves.  In case clock $A$ is very far away from clock $B$, the measurement probability tends to steady, which means that the probability of clock synchronization is not affected by gravitational interaction in this situation. We also present how the precision of clock synchronization is affected by the the gravitational interaction. The precision for the estimation of time difference depends to energy gaps and  the improvement of precision in quantum metrology is in fact an indicator for the  generation of quantum correlations.  In addition, the  quantum Fisher information is found to be very discriminable for different distances  between two clocks and it doesn't have a significant change for larger distances. 

\begin{acknowledgments}
This work is supported by the Science and Technology Planning Project of Hunan Province under Grant No. 2018RS3061;  and  the  Natural Science Fund  of Hunan Province  under Grant No. 2018JJ1016; and the National Natural Science Foundation
of China under Grant  No. 11675052 and No. 11475061.			
\end{acknowledgments}

\end{document}